\documentclass[10pt]{article}

\usepackage[dvips]{graphicx}
\usepackage{amsmath}
\usepackage{amssymb}

\begin{document}

\begin{center}

\vspace{0.5cm}
\textbf{\Large 
A simple and efficient numerical scheme to integrate non-local potentials}

\vspace{5mm} {\large N.~Michel}

\vspace{3mm}
\textit{CEA, Centre de Saclay, IRFU/Service de Physique Nucl{\'e}aire, F-91191 Gif sur Yvette, France }
\end{center}

\vspace{5mm}

\hrule \vspace{5mm} \noindent{\Large \bf Abstract}
\vspace*{5mm}

As nuclear wave functions have to obey the Pauli principle, potentials issued from reaction theory 
          or Hartree-Fock formalism using finite-range interactions contain a non-local part. Written in coordinate
           space representation, the Schr{\"o}dinger equation becomes
           integro-differential, which is difficult to solve,
           contrary to the case of local potentials, where it is an ordinary differential equation. A simple and powerful method has
           been proposed several years ago, with the trivially equivalent potential method, where non-local potential
           is replaced by an equivalent local potential, which is state-dependent and has to be determined iteratively.
           Its main disadvantage, however, is the appearance of divergences in potentials if the wave functions have nodes, 
           which is generally the case. We will show that divergences can be removed by a slight modification of the 
           trivially equivalent potential method, leading to a very
           simple, stable and precise numerical technique to deal
           with non-local potentials. Examples will be provided with the calculation of the Hartree-Fock potential
           and associated wave functions of $^{16}$O using the finite-range N$^3$LO realistic interaction.

\vspace{5mm} \noindent{\bf PACS}\\
{02.60.Nm} Integral and integrodifferential equations \\
       {03.65.Ge} Solutions of wave equations: bound states \\
       {03.65.Nk} Scattering theory \\
       {21.60.Jz} Nuclear Density Functional Theory and extensions
\section{Introduction}
\label{intro}
As protons and neutrons are indistinguishable in nuclei, Pauli principle must be taken into account in order to build
nuclear wave functions. This has the well-known consequence of the appearance of exchange potentials 
in the one-body Schr{\"o}dinger equation, in the optical potential of reaction theory \cite{Hodgson} 
or in the Hartree-Fock (HF) potential, issued from the variational principle applied to an antisymmetric wave function
of independent particles \cite{Messiah}. Exchange potentials are non-local, i.e~they are integral operators acting 
on the wave function at each point of space, so that Schr{\"o}dinger equation written in coordinate space representation
becomes an integro-differential equation. This type of equation cannot be handled by standard numerical methods used
in ordinary differential equations, such as midpoint and Henrici schemes \cite{Numerical_Recipes}.

Many methods have been devised to deal with non-local potentials. 
While some of them consider the implementation of bound states only \cite{Hooverman}, 
scattering states can be handled with the methods of Refs.\cite{Collins_Schneider,Gonzales,Kang_Chong},
where the Schr{\"o}dinger equation is transformed to an integral
equation.
Solving the Schr{\"o}dinger equation represented in momentum space has
also been considered (see Refs.\cite{Mondragon,Hagen}), 
as it becomes an integral equation therein as well.
However, in practice, momentum space has to be discretized \cite{Hagen}, so that wave functions do not
have proper asymptotic behavior when back-transformed to coordinate space. Moreover, consideration of
Coulomb potential cannot be handled exactly, its Fourier-Bessel function
being undefined due to its infinite range.
In the FRESCO reaction code (see Ref.\cite{Fresco} for numerical methods
employed therein), dealing with coupled-channel sets of equations involving non-local couplings, 
non-local parts in equations are replaced by source terms, converging iteratively to their
exact values, which is the most straightforward method to transform a
non-local equation into a local equation
(we will mention it from now on as the source method).
However, the source method usually converges very slowly, and sometimes even diverges. 
Indeed, even in the case of reaction theory, where local potential is dominant,
Pad{\'e} approximants have to be used in the FRESCO code in order to avoid instabilities \cite{Fresco}.

A very simple method has been introduced in
Ref.\cite{Vautherin_Veneroni}, where the non-local potential is replaced
by a state-dependent potential, the so-called trivially equivalent local potential (TELP),
and where the problem has to be solved iteratively as well. 
This scheme is particularly interesting in HF framework, 
where iterative scheme is the only one available as HF equations
are non-linear. As all considered equations become local, standard numerical methods for ordinary differential equations
can be used, so that the implementation of both bound and scattering states poses no problem.
Moreover, it is very stable in practice, contrary to the source method, so that it is widely used in nuclear theory
\cite{HFB_Grasso,PRC_Lithium,Rev_Mex_Fis}.
However, as the TELP method involves a division by the considered wave functions,
it contains poles, which are difficult to handle numerically. The solution proposed in Ref.\cite{Vautherin_Veneroni} 
consists in an interpolation of the potential near its poles, replacing the potential by a straight line connecting
two points before and after the pole, the distance between the two points going to zero iteration after iteration.
However, this procedure is difficult to apply, as there is no precise criterion to determine how fast the mentioned
distance has to decrease on the one hand, and, on the other hand, potentials become very large close to poles, which
can generate numerical inaccuracies. It would then be interesting to have the advantages of both source 
and TELP methods, i.e.~fast convergence of the iterative scheme and absence of divergences in potentials.

\section{Combination of source and TELP methods}
\label{sec:1}

\subsection{Local equivalent equation} \label{subsec}
For simplicity, we will consider spherically symmetric potentials only, even though the method can be readily extended
to more complicated situations, such as coupled-channel sets of equations. The Schr{\"o}dinger equation then reads:
\begin{eqnarray}
u''(r) &=& \left[ \frac{\ell(\ell+1)}{r^2} + \frac{2m}{\hbar^2} \left(v(r) + \frac{Z_\nu~C_c}{r} -
	    e \right) \right] u(r) \nonumber \\
&+& \frac{2m}{\hbar^2} \int_{0}^{+\infty} w(r,r') u(r')~dr' \label{Sch_eq},
\end{eqnarray}
where $Z_\nu$ is the charge of the target (proton) or is equal to zero
(neutron), $C_c$ is the Coulomb constant $\simeq 1.44$ MeV fm, $m$ is the effective mass of
the nucleon in MeV c$^{-2}$ units,
$u(r)$ is the calculated wave function, of orbital momentum $\ell$, and
energy $e$ in MeV units, 
$v(r)$ is the remaining local part of the potential in MeV units
and $w(r,r')$ its non-local part in MeV fm$^{-1}$ units. $v(r)$ and $w(r,r')$ are assumed to decrease quickly at large distance,
so that $u(r)$ becomes a linear combination of Coulomb wave functions for $r > R$, with $R$ sufficiently large.

We will now define the equivalent local equation used in the proposed method:
\begin{eqnarray}
u''(r) &=& \left[ \frac{\ell(\ell+1)}{r^2} 
+ \frac{2m}{\hbar^2} \left( \frac{}{} v_{loc}(r) - e \right) \right] u(r) 
+ \frac{2m}{\hbar^2} s(r) \label{Sch_eq_mod}, \\
v_{loc}(r) &=& v(r) + \frac{Z_\nu ~ C_c}{r} 
+ \frac{1 - F(r)}{u_{bef}(r)} \int_{0}^{+\infty} w(r,r') u_{bef}(r')~dr' \label{v_loc_def}, \\
s(r) &=& \frac{F(r)}{C_{bef}} \int_{0}^{+\infty} w(r,r') u_{bef}(r')~dr' \label{source_def}, \\
F(r) &=& \exp \left( -100 \left| \frac{u_{bef}(r)}{u'_{bef}(r)}
			  \right|^2 \right) 
 \left[ 1 - \exp \left( -100 \left| \frac{C_{bef}~r^{\ell + 1}}{u(r)} - 1 \right|^2 \right) \right], \label{F_def}
\end{eqnarray}
where $u_{bef}(r)$ is the normalized wave function obtained at previous iteration,
$v_{loc}(r)$ is the state-dependent local potential used in our method, $s(r)$ is a source term, 
$F(r)$ is a smoothing function, detailed afterward, and $C_{bef}$ is a normalization constant defined 
so that $u_{bef}(r) \sim C_{bef}~r^{\ell+1}$ for $r \sim 0$. 
Boundary conditions demanded for $u(r)$ in Eq.(\ref{Sch_eq_mod}) are $u(r) \sim r^{\ell+1}$ for $r \sim 0$,
and outgoing wave function condition at $r \rightarrow +\infty$ if one
considers a bound or resonant (Gamow) state \cite{Gamow,Gurney_Condon}.
The boundary condition for $u(r)$ at $r \sim 0$ implies the presence of $C_{bef}$ in Eq.(\ref{source_def}),
because, during integration of Eq.(\ref{Sch_eq_mod}), $u(r)$ is not
normalized whereas $u_{bef}(r)$ is. $u(r)$ will naturally be normalized
at the end of the calculation.
If $u(r)$ is a scattering state, no boundary condition at $r \rightarrow +\infty$ is required.
One can easily check that Eq.(\ref{Sch_eq_mod}) is equivalent to Eq.(\ref{Sch_eq}) if $u_{bef}(r) \propto u(r)$, 
hence at convergence of the iterative process. 

\subsection{Interest of the method}
The method embodied by Eqs.(\ref{Sch_eq_mod},\ref{v_loc_def},\ref{source_def},\ref{F_def}) is very close
to the TELP method of Ref.\cite{Vautherin_Veneroni}. 
TELP method is indeed recovered if one arbitrarily sets $F(r) = 0$ 
in Eqs.(\ref{v_loc_def},\ref{source_def}).
The presence of $F(r)$ is hence directly related to the zeroes of the function $u(r)$, 
which are the cause of the divergences in the TELP method.
$F(r)$ consists in two factors in Eq.(\ref{F_def}).
The first one is readily seen to be virtually zero except in the vicinity of the nodes
of $u(r)$, where it behaves like a Gaussian of maximal value equal to one, 
so that it cancels the divergences present in the TELP.
The second factor originates from the fact
that the TELP is well behaved at $r \sim 0$, even though $u(0) = 0$ \cite{Vautherin_Veneroni}. 
Thus, it is numerically more stable to have $v_{loc}(r)$ equal to the TELP close to $r = 0$. 
As the second factor is virtually equal to zero close to $r=0$, whereas it is otherwise almost equal to one,
it removes the action of the first factor at $r=0$, while leaving it unchanged for the other nodes of $u(r)$.
The used decay constants equal to 100 are rather arbitrary
and were empirically determined in order to have stable calculations for a large set of nuclei. 
As a consequence, $v_{loc}(r)$ is always finite close to the nodes of $u(r)$. Hence, the integration of Eq.(\ref{Sch_eq_mod})
is always numerically stable. Moreover, as $s(r)$ is non zero only in very small regions of considered radii,
it does not hinder convergence as in the source method. 
Note that $u'(r)$ does not enter Eq.(\ref{Sch_eq_mod}), 
so that Numerov and Henrici methods can be applied to solve it.
Consequently, both advantages of the source and TELP methods are present in the proposed numerical scheme.
Another interesting feature is the possibility to obtain rapidly approximate energies of bound and narrow resonant states
of Eq.(\ref{Sch_eq_mod}). For that, one diagonalizes $v_{loc}(r)$
with a basis of harmonic oscillator (HO) states. 
Due to the overall smallness of $s(r)$, eigenvalues of the HO matrix are very good starting energies, 
which are refined afterward with Newton method in order to determine the exact eigen-energies of Eq.(\ref{Sch_eq_mod}).

\section{Example: HF potential of $^{16}$O}
\label{sec:2}

\subsection{Motivation}
In order to illustrate the effectiveness of the technique described in Sec.(\ref{sec:1}),
we will consider the evaluation of the HF potential of $^{16}$O 
generated by the realistic interaction N$^3$LO \cite{N3LO}, 
renormalized within the low-momentum interaction framework \cite{Vlow_k}.
The maximal momentum in two-body relative space used for the latter is $\Lambda = 1.9$ fm$^{-1}$. 
Note that we do not aim at describing properly $^{16}$O properties at HF level, 
but simply at illustrating the proposed numerical scheme.
HF equations are solved iteratively using linear mixing method for HF potential
and a Woods-Saxon potential was used as starting point for the HF iterative process.
As we consider a rather small nucleus, a very large potential mixing of 80 \%
could be used, which resulted in a very quick convergence to the HF
solution in 32 iterations. If pairing interaction is used, in
the context of Hartree-Fock-Bogolyubov (HFB), linear mixing might be
insufficient, so that the more powerful modified Broyden method should
be used (see Ref.\cite{Broyden} for recent application to HFB formalism
and comparison to linear mixing method).
Moreover, as the N$^3$LO interaction is decomposed in a HO basis \cite{PRC_N3LO}, the non-local part of the
HF potential is a sum of separable functions of the form $f(r)~g(r')$. 
As convergence of shell model energies and eigenvectors is very quick with the number of HO basis states,
which was shown in Ref.\cite{PRC_N3LO}, 9 HO states per partial wave
are used in the decomposition of the N$^3$LO interaction in the present
work. The used HO parameter was $b = 2$ fm.
Hence, the integrations involving $w(r,r')$ in Eqs.(\ref{v_loc_def},\ref{source_def}) 
for each $r$ are replaced by sums over a few number of HO states, rendering wave functions determination very fast 
even though non-local operators are used. Note nevertheless that potential separability is not demanded in our method,
this property being used only to improve efficiency of calculations.

\subsection{Results}
\begin{figure*}
\resizebox{0.75\textwidth}{!}{\includegraphics{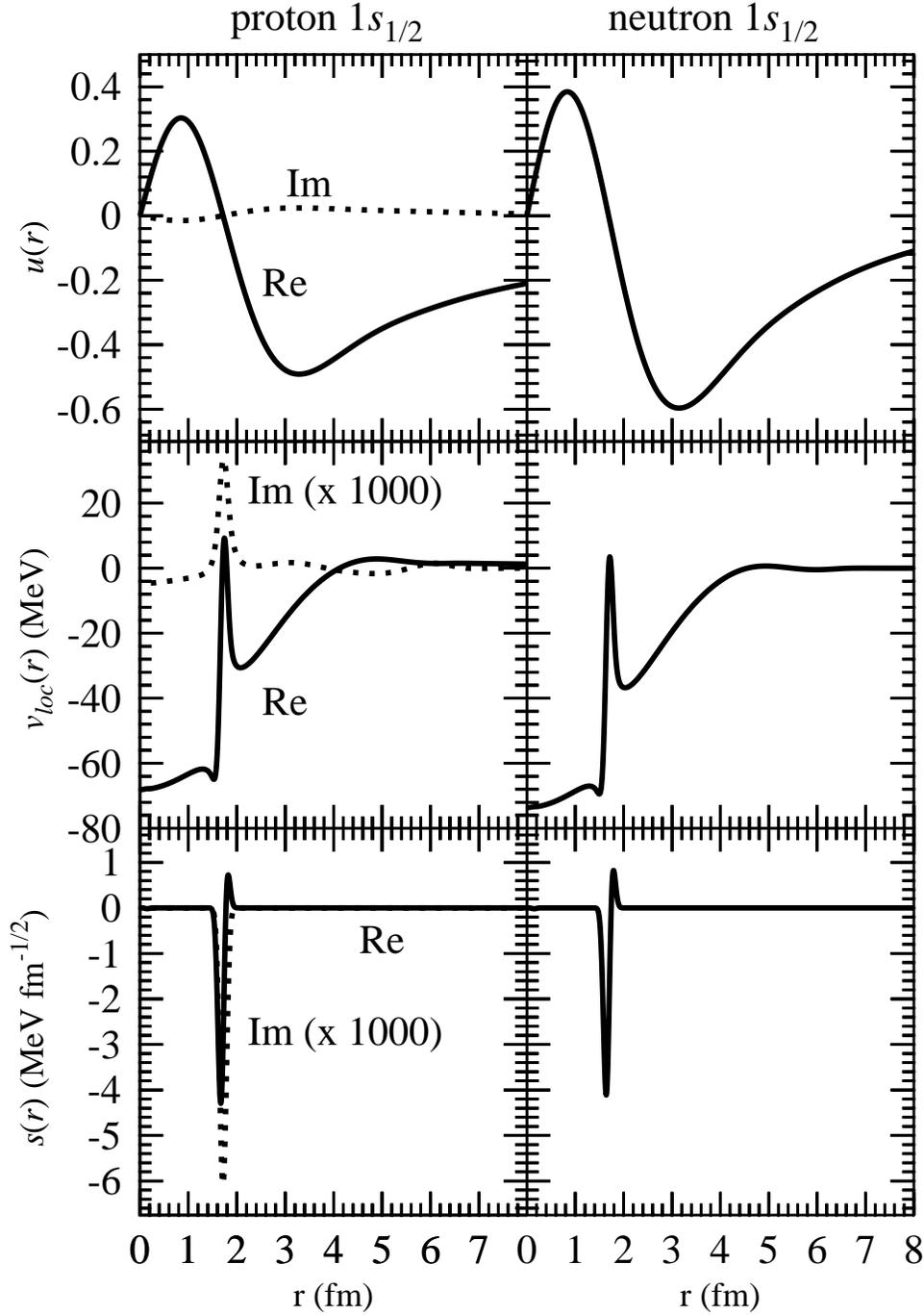}}
\caption{Wave functions (up), potentials (middle) and sources (bottom) of proton and neutron
 $1s_{1/2}$ HF states.
Real part (Re) is depicted as full line and imaginary part (Im) as dotted line.
Imaginary part has been multiplied by a factor of 1000 in the middle/left and
 bottom/left quadrants to be visible on the figure.
Imaginary parts are present for the proton case as proton HF state is
 resonant, whereas the bound neutron state is real, so that no imaginary
 part occurs therein. See Sec.(\ref{subsec}) for definitions of
 potentials and sources.}
\label{fig:1}
\end{figure*}

\begin{figure*}
\resizebox{0.75\textwidth}{!}{\includegraphics{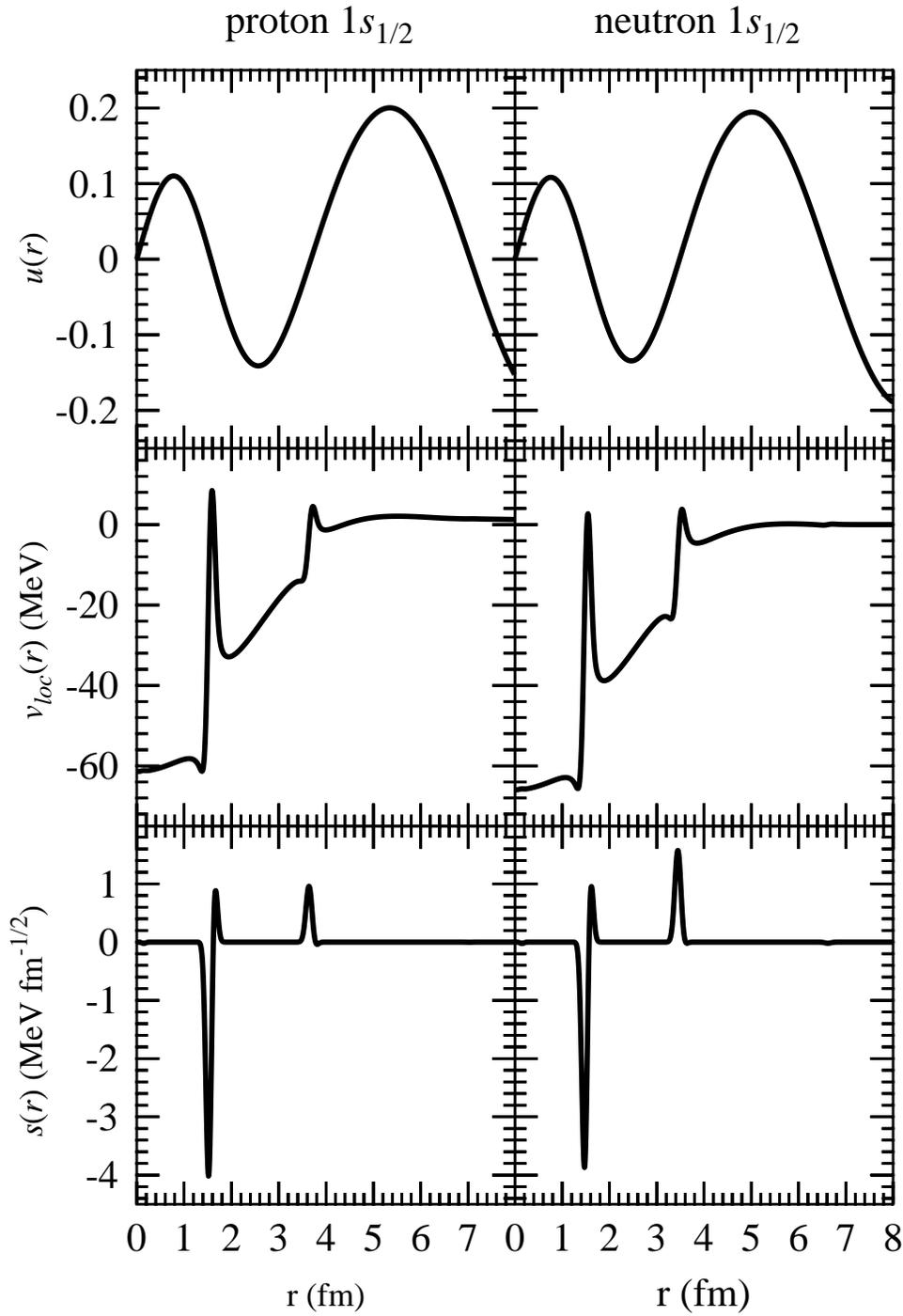}}
\caption{Same as Fig.(\ref{fig:1}), but with scattering proton and
 neutron $s_{1/2}$ states of momentum $k = 0.977$ fm$^{-1}$.}
\label{fig:2}
\end{figure*}

Energies of occupied single particle states and of unoccupied $1s_{1/2}$ states 
are shown in Tab.(\ref{tab:1}), where one can see that the proton $1s_{1/2}$ state is resonant. 
We will now concentrate on the proton and neutron $s_{1/2}$ states, as they contain nodes in the nuclear region,
which would induce divergences in TELP. Besides bound/resonant $s_{1/2}$
states, 30 scattering $s_{1/2}$ proton and neutron states have been calculated 
(see Tab.(\ref{tab:2}) for the value of their linear momenta). As the proton $1s_{1/2}$ state is resonant,
proton $s_{1/2}$ scattering states have been chosen to belong to a complex contour in $k$-space, as would be the case
in a Gamow Shell Model calculation
\cite{PRC_Lithium,Rev_Mex_Fis,PRL_GSM,PRC1_GSM,JPG_GSM}, 
where Berggren bases of complex energy-states consist in bound states, resonant states,
and contours of complex scattering states enclosing resonances \cite{Berggren}. Potentials $v_{loc}(r)$, sources $s(r)$ 
and wave functions $u(r)$ are illustrated in Fig.(\ref{fig:1}) for bound
neutron $1s_{1/2}$ state and resonant proton $1s_{1/2}$ state, and in Fig.(\ref{fig:2}) for two $s_{1/2}$
proton and neutron scattering states of linear momentum $k = 0.977$ fm$^{-1}$.
One can see that potential $v_{loc}(r)$ and source $s(r)$ vary very much close to the nodes of associated wave functions, 
but their maximal values in modulus remain sufficiently small not to generate numerical inaccuracies. Moreover, these variations
partially cancel in Eq.(\ref{Sch_eq_mod}), as $v(r)$ and $w(r,r')$ potentials are smooth in Eq.(\ref{Sch_eq}).
In order to check the accuracy of obtained wave functions, we have
calculated the overlaps between the considered $s_{1/2}$ states, in both proton
and neutron cases.
As one deals with unbound states, complex scaling method \cite{PRC1_GSM} is used to calculate radial integrals, which
diverge on the real axis. Average and maximal norms (used norm is $\max(|\Re(z)|,|\Im(z)|)$) of overlaps are shown in Tab.(\ref{tab:3}). As average values 
of overlaps are of the order of $10^{-9}$, calculated states are numerically orthogonal, 
which shows that the method described in Sec.(\ref{sec:1}) is very precise.

\begin{table}
 \caption{Single particle energies of occupied $0s_{1/2}$, $0p_{3/2}$
 and $0p_{1/2}$, and unoccupied $1s_{1/2}$ proton $(E_p)$ and neutron $(E_n)$
 HF states of $^{16}$O. They are given in MeV units. For the
 resonant $1s_{1/2}$ proton state, width is written between parentheses in
 keV units after its energy value.}
\label{tab:1}       
\begin{tabular}{lll}
\hline\noalign{\smallskip}
state & $E_p$ & $E_n$  \\
\noalign{\smallskip}\hline\noalign{\smallskip}
$0s_{1/2}$ & -64.657 & -69.444 \\
$0p_{3/2}$ & -29.221 & -33.821 \\
$0p_{1/2}$ & -19.753 & -24.065 \\
$1s_{1/2}$ & 0.519 (22.928) & -2.826 \\
\noalign{\smallskip}\hline
\end{tabular}
\end{table}

\begin{table}
 \caption{Linear momenta of scattering $s_{1/2}$ proton ($k_p$) and
 neutron ($k_n$) HF states in fm$^{-1}$. Used principal quantum number
 is arbitrary.}
\label{tab:2}
\begin{tabular}{lll}
\hline\noalign{\smallskip}
state & $k_p$ & $k_n$  \\
\noalign{\smallskip}\hline\noalign{\smallskip}
$2s_{1/2}$  & 0.105-i0.00469 & 0.105 \\
$3s_{1/2}$  & 0.123-i0.0231  & 0.123 \\
$4s_{1/2}$  & 0.15-i0.05  & 0.15 \\
$5s_{1/2}$  & 0.177-i0.0769  & 0.177 \\
$6s_{1/2}$  & 0.195-i0.0953  & 0.195 \\
$7s_{1/2}$  & 0.214-i0.0953  & 0.214 \\
$8s_{1/2}$  & 0.269-i0.0769  & 0.269 \\
$9s_{1/2}$  & 0.35-i0.05  & 0.35 \\
$10s_{1/2}$ & 0.431-i0.0231  & 0.431 \\
$11s_{1/2}$ & 0.486-i0.00469  & 0.486 \\
$12s_{1/2}$ & 0.523  & 0.523 \\
$13s_{1/2}$ & 0.616  & 0.616 \\
$14s_{1/2}$ & 0.75  & 0.75 \\
$15s_{1/2}$ & 0.885  & 0.885 \\
$16s_{1/2}$ & 0.977  & 0.977 \\
\noalign{\smallskip}\hline
\end{tabular}
\end{table}

\begin{table}
\caption{Maximal and average norms of the overlaps between the considered $s_{1/2}$ 
states, i.e.~ bound/resonant $0s_{1/2}$ and $1s_{1/2}$ states, and the scattering states described in Tab.(\ref{tab:2}). 
Overlaps are calculated with the complex scaling method \cite{PRC1_GSM}.}
\label{tab:3}
\begin{tabular}{lll}
\hline\noalign{\smallskip}
 & Maximal & Average \\
\noalign{\smallskip}\hline\noalign{\smallskip}
Proton  & 1.526 $\times 10^{-8}$ & 2.207 $\times 10^{-9}$ \\
Neutron & 1.287 $\times 10^{-8}$ & 1.307 $\times 10^{-9}$ \\
\noalign{\smallskip}\hline
\end{tabular}
\end{table}

\section{Conclusion}
Numerical methods allowing to solve Schr{\"o}dinger equations with non-local potentials are usually 
complicated or restricted to a given class of subproblems, 
such as the consideration of bound states uniquely. Source and TELP methods, on the contrary,
are very simple to code, but at the price of slow convergence and reduced stability for the former, and of the
appearance of potential divergences in the latter. 
The proposed method, by combining advantages of both source and TELP methods, allows to solve 
non-local Schr{\"o}dinger equation very quickly and precisely, which has been shown with the example of HF calculation
of $^{16}$O with finite-range realistic interaction N$^3$LO. 
Due to its simplicity and efficiency, it is a very interesting method to deal with the integration of 
non-local Schr{\"o}dinger equation.

\end{document}